\documentclass[12pt,preprint]{article}

\newcommand\simlt{\lower.5ex\hbox{$\; \buildrel < \over \sim \;$}}
\newcommand\simgt{\lower.5ex\hbox{$\; \buildrel > \over \sim \;$}}

\begin{document}
\title{Ultrahigh Energy Activity in Giant  Magnetar Outbursts }

\author{David Eichler
\footnote{Physics Department, Ben-Gurion University, Beer-Sheva
84105, Israel; eichler@bgumail.bgu.ac.il}} \maketitle
%
\begin{abstract}
The recent superflare of 27 December 2004 from the magnetar SGR
1806-20 was the brightest extrasolar flash ever recorded in the
modern era. The chances for seeing exotic ultrahigh energy (UHE)
radiation - neutrons, neutrinos, gamma rays and charged cosmic
rays - from it are far better from an energetic point of view than
from cosmological gamma ray bursts (GRBs). The chances for
detecting the various components are discussed in light of recent
data from the 27 December event.
\end{abstract}

\section{Introduction}
Giant flares from magnetars - most recently observed from
SGR1900+14 on 27 August 1998 and from SGR1806-20 on 27 December
2004 - were considered as a source of high energetic neutral
particles: neutrons, neutrinos, and photons (Eichler 2003, 2004,
Gelfand et al. 2005, Ioka et al. 2005, Halzen et al. 2005). Until
recently, their low occurrence rate in the galaxy, once every 20
years or so, rendered the subject as being of no obvious urgency.
Moreover, the fluxes of neutrons and neutrinos were scant enough
for the 27 August flux levels that a positive signal might have
been expected only very rarely. Much larger flares ($\ge 10^{46}$
ergs) were also considered in Eichler (2002) but the expected
occurrence rate in our galaxy, given the limits imposed by the
rate of short gamma ray bursts (GRBs), suggested that the event
rate per galaxy was at most one in several hundred years. This
estimate can be made without reference to extragalactic data by
assuming a dozen magnetars per galaxy and that each one has an
active phase lasting 3,000 to 10,000 years. At present, the
distance to which the December 27th flare could have been seen has
been estimated to be $50d_{15Kpc}$ Mpc (Palmer et al. 2005). Here,
$15d_{15Kpc}$ Kpc is the actual distance to SGR 1806-20 (Corbel
and Eikenberry 2004). This estimate, assuming a distance of 15
Kpc, suggests that the 27 December flare could have be seen in the
30,000 closest galaxies or so. The upper limit on detection of
such events is currently estimated to be about 50 per year
(Gehrels, private communication) so the upper limit on the event
rate per galaxy is approximately one per 600 years, with obvious
uncertainties due to small number statistics. The odds of it
happening once in 40 years per galaxy are thus small, though not
implausibly so.

The recent supergiant flare of 27 December
2004, which came as a surprise, can be attributed to luck. It may
be that our galaxy is somewhat atypical so that the a priori odds
of its happening once in a human lifetime are not implausibly
small; for example, not many galaxies the size of ours have
recently collided with a galaxy as large as the Magellenic Clouds
and the magnetar production rate in our galaxy could plausibly be
a factor of several times higher than in the average galaxy.
Magnetar production per unit mass in our galaxy and in the
Magellenic Clouds could be somewhat higher than average. This fact
could prove to be important in considering giant flares as a
source of UHE cosmic rays, which require an output of about
$10^{44}$ ergs per year (in UHE cosmic rays) per average large
galaxy.

In any event, the 27 December event, however improbable it may
have been, has nevertheless occurred and has refocused attention
on the issue of ultrahigh energy neutrals (Gelfand et al. 2005).
Because it was 100 times brighter than the next brightest event,
that of 27 August 1998, the fluxes that are worth considering
could have been quite detectable with existing underice neutrino
detectors (AMANDA) and airshower arrays. Since the event has
already happened, either these fluxes were detected or they
weren't. The purpose of this letter is to provide the
experimentalist with a general guideline of what might be
expected. We briefly review the prospects for ultrahigh energy
neutrons, neutrinos, photons, and charged cosmic rays, in light of
the data for the 27 December event that have been reported thus
far.

 Neutrons can be detected only at a Lorentz factor $\ge10^9$
(Eichler 2003), or else they would not cross the galaxy before
decaying. Neutrinos are best detected between energies of 1 and
100 TeV. Photons could have been detected by airshower arrays at
energies above $10^{14}$ eV and by MILAGRO above $10^{12}$ eV.
Charged cosmic rays can be seen only over the course of many years
as they do not propagate to earth in a straight line.

The immediate questions that arise upon which
theorists might be able to provide some guidance are how the
energy in the flare might distribute itself among these and other
forms of energy, what energy range is predicted, and what time
scale is predicted. These depend strongly on whether the flare
puts most of its energy into pair plasma, relatively devoid of
baryons, and how much goes into baryon-contaminated plasma, and at
what Lorentz factor these components are ejected. A second
question is the spectrum of ultrahigh energy charged particles
that are produced.

Given the basic hypothesis that the flare is
due to magnetic reconnection in the magnetosphere, one naively
expects that the plasma be highly pair-dominated, because baryons
are constrained by gravity not to populate the magnetosphere.
However, there may be several reasons to expect baryons: 1)
Baryons in some small measure may populate the magnetosphere due
to electromagnetic forces, especially if they are mobilized as
carriers of the electric currents. Such a population of dilute
baryons can none the less absorb significant amounts of energy if
they dominate the inertia of the magnetically reconnecting plasma
(Eichler 2003), and they would attain particularly dramatic
individual energies. 2) Baryons may be dredged up from just below
the surface of the magnetar. In a recent paper (Gelfand et al.
2005) it has been shown that more than $10^{24}$ g of material is present in the radio afterglow nebula and the natural explanation for this is that they were ejected from the magnetar itself. Alternatively, the baryons may have been in the ambient medium but
the measured expansion trajectory is difficult to explain unless
these baryons were concentrated improbably close to the magnetar
(Granot et al. 2005).

The expanding radio nebula has been fit with a baryonic shell of
mass greater than $10^{24}$ g and expansion velocity of about
0.3c, assuming that it is roughly spherical (Gelfand et al. 2005).
This does not preclude a more relativistic component, though such
a relativistic component would probably be required to store its
energy in relativistic baryons in order to avoid annihilation.
However, in another paper from this collaboration, Granot et al.
(2005) argue that the radio emission that is currently being
observed cannot be attributed to relativistic outflow, and this
constraint sets a rather high minimum for the amount of rest mass.
If the ejecta are one-sided, the inferred expansion velocity is
about 0.6c but the opening angle is considerably smaller than
$4\pi$, the energy estimates as expressed by Gelfand et al. (2005)
are thus insensitive to this matter. There exists the possibility
that this mass was actually ambient material, as the magnetar
probably lives in a molecular cloud complex. However, the energy
transfer from relativistic ejecta to the ambient material is
unlikely to be 100 percent efficient in view of Rayleigh-Taylor
instabilities and this would only raise the minimum energy
requirements on the relativistic ejecta, particularly if the
relativistic ejecta in their present state are mostly one-sided.
The opacity of matter leaving the magnetar surface increases with
distance as the magnetic field decreases and it is not unlikely
that a large fraction of the ejected energy, even if originally in
the form of radiation, could drive some of the ejected matter to
relativistic velocities and end up in that form.

 To summarize all of the above, the total flare energy, at least
$2 \times 10^{46}$ ergs can distribute itself in extremely high
Lorentz factor plasma, relativistic baryons, and non-relativistic
baryons, though it seems that the component in relativistic
baryons is energetically the smallest of the three. The present
radio nebula was driven by non-relativistic baryons (Gaensler et
al. 2005, Gelfand et al. 2005, Granot et al. 2005), so we now have
observational evidence for at least this component, as well as the
gamma ray component, which almost certainly came from baryon-poor
pair plasma.

\section{Mass Loss}
The fits to the expanding radio nebula suggest a minimum mass
ejection of several times $10^{24}$ g. There are several reasons
for believing this mass was ejected during the initial hard spike
phase of the giant flare. The tail phase, which lasted several
minutes, has a well-defined time structure (on a sub-second time
scale) suggesting that the optical depth through which it was
observed was small, yet this is unlikely to have been the case if
the baryonic outflow had been emitted over many rotation periods
(as it would demand a line of sight that was much cleaner than
average).
If, on the other hand, the matter was ejected
during the first 0.5 seconds of the flare along many lines of
sight, none of them need have swept across our line of sight
during such a small fraction of the rotation period (7.6 s).
Although there was apparently more than enough matter in the
ejecta to have obscured the observed gamma ray flare if it were
present at the source, we suggest that it was emitted extremely
anisotropically. Probably, the baryon content of the energy
outflow was very strongly dependent on its point of origin and/or
time of origin at the magnetar surface. This strong dependence was
accomplished without an event horizon and we suggest that it is an
intrinsic property of magnetic field annihilation in a highly
stratified medium. The extent of mass loading on
magnetically-driven outflow can be very sensitive to initial
conditions. In a purely one-dimensional situation with the
magnetic field perpendicular to the gravitational force, the mass
density $\rho(t)$ on any given field line is proportional to
$\rho(0) B(t)/B(0)$. In hydrostatic equilibrium
\begin{equation}
\rho g = \rho(0) B/B(0)g = \frac{d}{dz} (B^2/8 \pi) = B
\frac{d}{dz} (B/4 \pi)
\end{equation}
Writing g as $\frac{d\Phi}{dz}$, this can be integrated to
\begin{equation}
B(0) - B(z) = 4\pi \int \rho(0)/B(0) d\Phi
\end{equation}

Given that, in a realistic geometry, the right hand-side is bounded, the change in field
with altitude, depicted by the left-hand side, is also bounded.
This shows that depending on the initial mass loading, some field
lines will blow off to infinity and others will remain bound to
the magnetar. However, in a more realistic geometry, magnetic
fields would arch up and matter would fall away from the rising
apex of the arch back towards the star and the mass loading near
the apex could decrease substantially with time. Thus magnetic
fields rising from below the surface could shed their matter if
they rose sufficiently gradually. Moreover, much of the
reconnection could take place above the surface at an altitude
that guaranteed a nearly vanishing baryon density. On the basis of
the above considerations, we conclude that the amount of matter
dragged out by erupting magnetic field lines is not only hard to
predict, but is likely to be highly variable as a function of
space and time on the surface. This squares with the naively
paradoxical observations that seem to indicate both a large gamma
ray flux directly from the surface as well as a mass loss. A
similar explanation might be given for the coexistence of prompt
gamma rays and after-glow-generating ejecta from cosmological
distant gamma ray bursts (GRBs). Note, however, that in the case
of cosmological GRB, the duration of the event is much longer than
the rotation period of the central powerhouse, whereas in the case
of the initial hard spike phase of giant flares from SGRs, the
reverse is true. As rapid rotation is likely to blend baryon-rich
components and baryon-poor components on any given line of sight,
cosmological gamma ray bursts might exploit a systematic polar
angle-dependence in the baryon richness such as might be expected
from a black hole-accretion disk system (Levinson and Eichler
1993).

\section{Particle Acceleration}
We now address the question of particle acceleration. We consider
both shock acceleration and bulk acceleration following magnetic
reconnection.

Particles can be accelerated by internal shocks in relativistic
outflows to energies as high as $\sim 10^{21}B_{15}$ eV (e.g., Levinson and Eichler 1993), where B is the magnetic field strength at the base of the flow, here a
neutron star. (In this paper numerical subscripts obey the convention $Q_n=10^{-n}Q$ in cgs units unless otherwise stated.) This estimate was made for an
outflow from a rotating neutron star considering the potential
drop along open field lines and with due allowance for a reduction
in the highest possible energy for internal shocks. Strictly
speaking, a flow characterized by magnetic field B velocity $\beta
c$ and transverse radius R can accelerate particles of charge Ze through a
maximum energy of
\begin{equation}
E_{max}=\sigma m_p c^2 \equiv ZeBR\beta \label{emax}
\end{equation}
(Eichler 1981, the so-called Hillas limit). Here $m_p$ is the mass
of the proton, Ze is the charge of the particle, and $\beta c$ is
the shock velocity. For an outflow of $\beta c$ from a neutron
star of radius $R_{NS}$, $\sigma \sim 10^{14.5} \beta
B_{15}R_{NS}$. If the source is rapidly rotating so that B
decreases as 1/r, then assuming R and r are of the same order,
$\sigma$ remains roughly constant with r. In the case of outflow
from a magnetar, however, the flow out to nearly the light
cylinder is not strongly effected by rotation and the field lines
are probably radial. In this case, B decreases as $1/r^2$ and
hence $\sigma (r) \propto 1/r$. At a characteristic radius r of
$10^{10} r_{10}$ cm $\sigma$ is thus $10^{10.5}\beta
B_{15}/r_{10}$ and it thus difficult to accelerate protons beyond
$10^{20}$ eV if the flow stretches the magnetic field radially
to $10^{10}$ cm.

The maximum energy is, in any case, limited by ion-synchrotron
radiation to
\begin{equation}
E_{max}=70\sigma^{\frac{3}{5}}(\frac{m}{m_p})^{\frac{2}{5}}m_pc^2
B_{15}^{-\frac{1}{5}} \label{emax2}
\end{equation}
where m is the mass of the particle (Eichler 2003), so that close
to the neutron star it is likewise difficult to accelerate protons
beyond $10^{20}$ eV. Assuming that $B\propto 1/r^2$, it follows
that $\sigma\propto 1/r$ and $E_{max}$ decreases with r. So the
maximum value to which a proton can be accelerated is $6\times
10^{19}B_{15}^{\frac{2}{5}}R_{NS,6}^{\frac{3}{5}}$ eV. Heavier
ions could attain a higher total energy but would be limited to a
lower energy per nucleon, and it is unlikely that they would
survive the intense radiation field intact if accelerated close to
the surface.

Let us now consider the efficiency with which the highest energy
particles can be accelerated. If the shocks are relativistic,
models based on small angle scattering predict (Bednarz and
Ostrowski 1998, Vietri 2003, Keshet and Waxman 2004) that their
spectral index is -p = -2.25. This implies that the energy
component in neutrons at $\Gamma \ge 10^9$ is less than $10^{-2}$
of that in the shock accelerated particles. Large angle scattering
in relativistic shocks, on the other hand (Ellison and Double,
2002) gives rise to very hard spectra and eliminates this problem. It is
possible, of course, that subrelativistic shocks can be embedded
in a highly relativistic outflow. Subrelativistic shocks, if at a
high enough Mach number, can, in fact, put most of their energy
into particles at the highest energy.

Bulk acceleration following magnetic reconnection is still a
somewhat open question. Lyutikov and Uzdenski (2003) conjectured
that Lorentz factors as high as $\sigma$ are attained within the
reconnection region. Lyubarsky (2005) constructed a model in which
the Lorentz factor of the material ejected from the reconnection
region is always below $\sigma ^{1/2}$, with the maximum attained
only for reconnection of field lines that are anti-parallel.
Eichler (2003) made the starting assumption that $\Gamma$ is of
the order of $\sigma^{1/2}$ without proof, and considered the
consequences of turbulent ejecta with this typical value for
$\Gamma$. In ultrarelativistic turbulence, second order Fermi
acceleration can be extremely efficient and accelerate particles
up to the limits mentioned above. Theoretical steady-state solutions to force-free
electrodynamics typically suggest that for asymptotic outflows,
the Lorentz factor increases linearly with radius and that the
magnetosonic point (where $\Gamma \sim \sigma^{1/2}$ for an
extremely high $\gamma$ outflow) would occur only at very large
radii. (The force-free electrodynamic approximation, in any case,
breaks down near the fast magnetosonic point, where, by
definition, inertia is important.) At very
large radii, many of the emission mechanisms discussed in Eichler
(2003) would not be relevant. However, it is possible that in time-dependent explosive
outflows the Lorentz factor is higher much closer to the surface (Lyubarsky in preparation).
Close to the surface, protons easily generate neutrons and
neutrinos by photopion reactions (e.g., Eichler 1978). The neutrinos will typically
carry 5 to 10 percent of the proton energy and emerging neutrons
would contain much or most of the initial proton energy and
arrive essentially simultaneously with photons emitted at the same
place and time.

Now consider charged UHE cosmic rays. The discussion here presents
a less optimistic picture than that of Asano et al. (2005). A
particle of charge Ze and energy $10^8 E_8$ in the interstellar
medium (where $B\sim 3\mu$G) has a gyroradius $r_g$ of $6\times
10^{22}E_8B_{-5.5}$ cm. This is comparable to the distance from
us. In traversing $d=10^{22.5}d_{22.5}$ cm, its traversed distance
deviates from that of a straight line connecting its end-points by
$2r_g(\theta-sin\theta)$ where $2rsin\theta=d$. To lowest order in
$\theta$, this yields $d^3/16r_g^2$. Here we have assumed a
constant field. A more detailed model by Alvarez-Muniz, Engel and
Stanev (2002) yields a delay of only about $10^2B_{-5.5}/E_8$
years from a distance of 20 Kpc. This estimate is in any case
highly uncertain as our knowledge of the Galactic field is
limited. Turbulence can reduce the delay for some particles and
increase it for others, but the presence of an underlying large
scale field would suggest a minimum delay for most particles.
The fluence in UHE cosmic rays would be
$1\times10^2\epsilon_{UHE}/E_8$ particles per km$^2$. Here
$\epsilon_{UHE}$ is the efficiency with which UHE cosmic rays are
produced relative to the burst energy, $2\times10^{46}
d_{15Kpc}^2$. A liberal estimate for $\epsilon_{UHE}$ in any given
logarithmic interval of energy is about 10 percent (Ellison and
Eichler 1985). A liberal estimate for the enhanced flux of UHE
cosmic rays in the energy range of $3\times10^{19}$ - $10^{20}$ eV
from the direction of the Galactic center is thus of order
0.1/km$^2$-yr.
It is worth looking for weak anisotropies in the Galactic disk
from previous magnetar outbursts as their event rate in the Galaxy
is probably more than one per 1000 years. A flux of even $10^{-2}$
per km$^2$-yr at $E_8\simeq 1$ confined to 0.1 radian of the
Galactic plane could be detectible with AUGER, which should detect
a total of 100 per year at these energies.

The question of whether giant flares from magnetars could provide
all of the UHE cosmic ray background is not much changed by the
huge energetics of the 27 December event. A magnetar has of the
order of $10^{47}$ ergs to release, regardless of how this
quantity may be divided into individual bursts. In our Galaxy, the
production rate of known magnetars appears to be 1 to 3 per 1000
years. This follows directly from the fact that there are 12,
including those in the Magellanic Clouds and that their active
lifetime, as deduced from their association with supernova
remnants, appears to be several thousand years. This suggests that
magnetars have barely enough energy to account for the UHE cosmic
rays and, given the uncertainties both in the theory and
observations, little more can be said at the present time. Whether
our Galaxy is completely typical in its magnetar production rate
is also an open question at present. However, observations of
extra-galactic magnetar flares, which should be available from
Swift (Eichler 2002) should help settle this question.

\section{Summary of scenarios for UHE emission}

To summarize, it appears to this author that most of the
theoretical possibilities that have been, or are likely to be,
discussed in the literature are presently possible, and are even
supported by observations of the radio nebula. The fraction of
energy that is in subrelativistic baryons appears from the radio
data to be at least 1 percent of the flare energy, but any
energetically plausible higher value is also consistent with the
data. It is unlikely that the vast majority of the blast energy is
in ultrarelativistic baryons or pairs, as it would have produced a
more rapidly expanding nebular shell. However, the amount of
energy in such a component may easily be within an order of
magnitude or so of that in the subrelativistic baryons
(Ramirez-Ruiz, private communication) because it could have been
slowed and overtaken by the latter within the first ten days after
the explosion. The huge energy emitted in gamma rays is probably
even larger than the blast energy and is therefore even less
likely to be matched by a comparable component in
ultrarelativistic pairs or even ultrarelativistic baryons.
However, the present observations of the radio nebula probably
admit as much as 10 percent of the flare energy in
ultrarelativistic baryonic outflow. The observations imply at
least $10^{44}$ ergs and as much as $10^{46}$ ergs in modestly
relativistic baryons and admit as much as 10 percent of this
quantity in ultrarelativistic outflow. The Lorentz factor of such
outflow can be anywhere between 1.1 and $10^{14}$, and it is quite
reasonable to suppose that we will receive a diverse sample from
this wide range.

The prompt neutrons, which require $\Gamma\ge10^9$, could be
detected with high statistical significance if they are, in fact,
efficiently produced. UHE protons, whose arrival even at the
highest conceivable energies would be spread out at least several
thousand years, are less likely to be detected with overwhelming
statistical significance, but even a fluence of $10^{-2}$ per
km$^2$-year, comparable to the background flux, would be a
statistically significant signal in AUGER given its large area. These particles need not necessarily arise from SGR1806-20; they may arise from other galactic magnetars as well. While we have argued that the maximum energy of protons is
unlikely to exceed $10^{20}$ eV, high energy particles should,
nevertheless, be looked for.

Plausible values for the neutrino flux from the 27 December event
have been very recently discussed by Gelfand et al. (2005), Ioka
et al. (2005), and Halzen et al. (2005). There are large
uncertainties in the predicted flux but, given the evidence for
baryonic ejection (Gelfand et al., 2005) and the huge total
fluence at Earth, this event is arguably the most promising
transient source of neutrino to date.

Ultrahigh energy photons are easily produced given particle
acceleration, but whether they escape is problematic. Levinson and
Eichler (2000) have presented a detailed analytic calculation of
escape criteria for UHE cosmic rays. For $\gamma$-ray energy
spectrum of $E^{-2}$ (most of the spectra give less photon-photon
opacity) and a $\gamma$-ray energy of $\epsilon_{\gamma} m_ec^2$,
the gammaspheric radius is given by
\begin{equation}
r_{\gamma}(\epsilon_{\gamma})=2.8\times10^{10}\frac{L_{51}}{\Gamma_2^{4}}
\epsilon_{\gamma}\ {\rm cm}
\end{equation} where L is the $\gamma$-ray luminosity.
During the initial hard spike of the 27 December flare, which
lasted $\delta t\sim 0.3$ s, $L_{51}\sim
 10^{-4}$. Photons more energetic than $10^{14}$ eV have $\epsilon
 _{\gamma} \ge 2\times 10^8$, so in order for them to escape
  from within $\Gamma^2c\delta t$ of the source, the Lorentz factor of the
  outflow, $\Gamma$ would have to exceed $10^2$, and this threshold is only weakly dependent on $\epsilon_{\gamma}$. This is
  achievable by magnetic reconnection in baryon-poor regions of the
  magnetosphere, where bulk motions with Lorentz factors as high as $\sigma^{1/2}$
  are in principle possible at the reconnection site ( Eichler 2003, Lyubarsky
  2005) and perhaps as high as $\sigma$ in the post reconnection
  flow (Lyutikov and Uzdenski 2003).

  I thank Drs. B. Gaensler, N. Gehrels, Y. Gelfand, J. Granot, C.
Kouveliotou, Y. Lyubarsky, E. N.   Parker, E. Ramirez-Ruiz, and P.
Woods for useful discussions. This research was supported by the
Israel-US Binational Science Foundation, an Israel Science
Foundation Center of Excellence Award, and the Arnow Chair of
Theoretical Astrophysics.

\end{document}